\documentclass[tightenlines,superscriptaddress,10pt,twocolumn,aps,prd]{revtex4-1}

\usepackage{geometry}
\usepackage{graphicx,wrapfig}
\usepackage[export]{adjustbox}
\usepackage{epsfig}
\usepackage{float}
\usepackage{flafter}
\usepackage{subfigure}
\usepackage{sidecap}
\usepackage{notes2bib}
\usepackage{gensymb}
\usepackage{url}
\usepackage{amsmath}
\usepackage{latexsym}
\usepackage{amssymb}
\usepackage{xcolor}
\usepackage{color}
\usepackage{slashed}
\usepackage{units}
\usepackage{siunitx}
\usepackage{notoccite}
\usepackage[utf8]{inputenc}
\usepackage[english]{babel}
\usepackage{csquotes}
\usepackage{dcolumn}
\usepackage{bm}
\usepackage{color}
\usepackage{enumitem}
\usepackage{pifont}
\usepackage{textcomp}
\usepackage{tikz}
\usepackage{colortbl}
\usepackage[version=4]{mhchem}
\usepackage{xspace}
\usepackage{placeins}
\usepackage{tabularx}
\usepackage{hyperref}
\usepackage{amssymb}
\usepackage{hyperref}
\usepackage{lineno}

\usepackage{pgfgantt}
\usepackage[ampersand]{easylist}

\newcommand{\cevns}{CEvNS}

\begin{document}

\onecolumngrid

\title{Measurement of the Coherent Elastic Neutrino-Nucleus Scattering Cross Section on CsI by COHERENT}

\widowpenalty10000
\clubpenalty10000
\renewcommand\floatpagefraction{1}
\renewcommand\topfraction{1}
\renewcommand\bottomfraction{1}
\renewcommand\textfraction{0}

\setlength{\belowcaptionskip}{-10pt} 

\renewcommand{\thesection}{\arabic{section}}
\renewcommand{\thesubsection}{\thesection.\arabic{subsection}}
\renewcommand{\thesubsubsection}{\thesubsection.\arabic{subsubsection}}

\makeatletter
\renewcommand{\p@subsection}{}
\renewcommand{\p@subsubsection}{}
\makeatother


\newcommand{\Mephidesc}{\affiliation{National Research Nuclear University MEPhI (Moscow Engineering Physics Institute), Moscow, 115409, Russian Federation}}
\newcommand{\Dukedesc}{\affiliation{Department of Physics, Duke University, Durham, NC, 27708, USA}}
\newcommand{\TUNLdesc}{\affiliation{Triangle Universities Nuclear Laboratory, Durham, NC, 27708, USA}}
\newcommand{\UTKdesc}{\affiliation{Department of Physics and Astronomy, University of Tennessee, Knoxville, TN, 37996, USA}}
\newcommand{\ITEPdesc}{\affiliation{Institute for Theoretical and Experimental Physics named by A.I. Alikhanov of National Research Centre ``Kurchatov Institute'', Moscow, 117218, Russian Federation}}
\newcommand{\ORNLdesc}{\affiliation{Oak Ridge National Laboratory, Oak Ridge, TN, 37831, USA}}
\newcommand{\USDdesc}{\affiliation{Physics Department, University of South Dakota, Vermillion, SD, 57069, USA}}
\newcommand{\NCSUdesc}{\affiliation{Department of Physics, North Carolina State University, Raleigh, NC, 27695, USA}}
\newcommand{\Sandiadesc}{\affiliation{Sandia National Laboratories, Livermore, CA, 94550, USA}}
\newcommand{\UWdesc}{\affiliation{Center for Experimental Nuclear Physics and Astrophysics \& Department of Physics, University of Washington, Seattle, WA, 98195, USA}}
\newcommand{\LANLdesc}{\affiliation{Los Alamos National Laboratory, Los Alamos, NM, 87545, USA}}
\newcommand{\Laurentiandesc}{\affiliation{Department of Physics, Laurentian University, Sudbury, Ontario, P3E 2C6, Canada}}
\newcommand{\CMUdesc}{\affiliation{Department of Physics, Carnegie Mellon University, Pittsburgh, PA, 15213, USA}}
\newcommand{\IUdesc}{\affiliation{Department of Physics, Indiana University, Bloomington, IN, 47405, USA}}
\newcommand{\VTdesc}{\affiliation{Center for Neutrino Physics, Virginia Tech, Blacksburg, VA, 24061, USA}}
\newcommand{\NCCUdesc}{\affiliation{Department of Mathematics and Physics, North Carolina Central University, Durham, NC, 27707, USA}}
\newcommand{\UFdesc}{\affiliation{Department of Physics, University of Florida, Gainesville, FL, 32611, USA}}
\newcommand{\Tuftsdesc}{\affiliation{Department of Physics and Astronomy, Tufts University, Medford, MA, 02155, USA}}
\newcommand{\SNUdesc}{\affiliation{Department of Physics and Astronomy, Seoul National University, Seoul, 08826, Korea}}
\author{D. Akimov}\Mephidesc
\author{P. An}\Dukedesc\TUNLdesc
\author{C. Awe}\Dukedesc\TUNLdesc
\author{P.S. Barbeau}\Dukedesc\TUNLdesc
\author{B. Becker}\UTKdesc
\author{V. Belov }\ITEPdesc\Mephidesc
\author{I. Bernardi}\UTKdesc
\author{M.A. Blackston}\ORNLdesc
\author{C. Bock}\USDdesc
\author{A. Bolozdynya}\Mephidesc
\author{J. Browning}\NCSUdesc
\author{B. Cabrera-Palmer}\Sandiadesc
\author{D. Chernyak}\altaffiliation[Now at: ]{Institute for Nuclear Research of NASU, Kyiv, 03028, Ukraine}\USDdesc\altaffiliation[Now at: ]{Institute for Nuclear Research of NASU, Kyiv, 03028, Ukraine}\altaffiliation[Now at: ]{Institute for Nuclear Research of NASU, Kyiv, 03028, Ukraine}
\author{E. Conley}\Dukedesc
\author{J. Daughhetee}\ORNLdesc
\author{J. Detwiler}\UWdesc
\author{K. Ding}\USDdesc
\author{M.R. Durand}\UWdesc
\author{Y. Efremenko}\UTKdesc\ORNLdesc
\author{S.R. Elliott}\LANLdesc
\author{L. Fabris}\ORNLdesc
\author{M. Febbraro}\ORNLdesc
\author{A. Gallo Rosso}\Laurentiandesc
\author{A. Galindo-Uribarri}\ORNLdesc\UTKdesc
\author{M.P. Green }\TUNLdesc\ORNLdesc\NCSUdesc
\author{M.R. Heath}\ORNLdesc
\author{S. Hedges}\Dukedesc\TUNLdesc
\author{D. Hoang}\CMUdesc
\author{M. Hughes}\IUdesc
\author{T. Johnson}\Dukedesc\TUNLdesc
\author{A. Khromov}\Mephidesc
\author{A. Konovalov}\Mephidesc\ITEPdesc
\author{E. Kozlova}\Mephidesc\ITEPdesc
\author{A. Kumpan}\Mephidesc
\author{L. Li}\Dukedesc\TUNLdesc
\author{J.M. Link}\VTdesc
\author{J. Liu}\USDdesc
\author{K. Mann}\NCSUdesc
\author{D.M. Markoff}\NCCUdesc\TUNLdesc
\author{J. Mastroberti}\IUdesc
\author{P.E. Mueller}\ORNLdesc
\author{J. Newby}\ORNLdesc
\author{D.S. Parno}\CMUdesc
\author{S.I. Penttila}\ORNLdesc
\author{D. Pershey}\Dukedesc
\author{R. Rapp}\CMUdesc
\author{H. Ray}\UFdesc
\author{J. Raybern}\Dukedesc
\author{O. Razuvaeva}\Mephidesc\ITEPdesc
\author{D. Reyna}\Sandiadesc
\author{G.C. Rich}\TUNLdesc
\author{J. Ross}\NCCUdesc\TUNLdesc
\author{D. Rudik}\Mephidesc
\author{J. Runge}\Dukedesc\TUNLdesc
\author{D.J. Salvat}\IUdesc
\author{A.M. Salyapongse}\CMUdesc
\author{K. Scholberg}\Dukedesc
\author{A. Shakirov}\Mephidesc
\author{G. Simakov}\Mephidesc\ITEPdesc
\author{G. Sinev}\altaffiliation[Now at: ]{South Dakota School of Mines and Technology, Rapid City, SD, 57701, USA}\Dukedesc\altaffiliation[Now at: ]{South Dakota School of Mines and Technology, Rapid City, SD, 57701, USA}
\author{W.M. Snow}\IUdesc
\author{V. Sosnovstsev}\Mephidesc
\author{B. Suh}\IUdesc
\author{R. Tayloe}\IUdesc
\author{K. Tellez-Giron-Flores}\VTdesc
\author{I. Tolstukhin}\altaffiliation[Now at: ]{Argonne National Laboratory, Argonne, IL, 60439, USA}\IUdesc\altaffiliation[Now at: ]{Argonne National Laboratory, Argonne, IL, 60439, USA}
\author{E. Ujah}\NCCUdesc\TUNLdesc
\author{J. Vanderwerp}\IUdesc
\author{R.L. Varner}\ORNLdesc
\author{C.J. Virtue}\Laurentiandesc
\author{G. Visser}\IUdesc
\author{T. Wongjirad}\Tuftsdesc
\author{Y.-R. Yen}\CMUdesc
\author{J. Yoo}\SNUdesc
\author{C.-H. Yu}\ORNLdesc
\author{J. Zettlemoyer}\altaffiliation[Now at: ]{Fermi National Accelerator Laboratory, Batavia, IL, 60510, USA}\IUdesc\altaffiliation[Now at: ]{Fermi National Accelerator Laboratory, Batavia, IL, 60510, USA}

\begin{abstract}
We measured the cross section of coherent elastic neutrino-nucleus scattering (\cevns{}) using a CsI[Na] scintillating crystal in a high flux of neutrinos produced at the Spallation Neutron Source (SNS) at Oak Ridge National Laboratory.  New data collected before detector decommissioning have more than doubled the dataset since the first observation of \cevns{}, achieved with this detector.  Systematic uncertainties have also been reduced with an updated quenching model, allowing for improved precision.  With these analysis improvements, the COHERENT collaboration determined the cross section to be $(165^{+30}_{-25})\times10^{-40}$~cm$^2$, consistent with the standard model, giving the most precise measurement of \cevns{} yet.  The timing structure of the neutrino beam has been exploited to compare the \cevns{} cross section from scattering of different neutrino flavors.  This result places leading constraints on neutrino non-standard interactions while testing lepton flavor universality and measures the weak mixing angle as $\sin^2\theta_{W}=0.220^{+0.028}_{-0.026}$ at $Q^2\approx(50\text{ MeV})^2$.  
\end{abstract}

\maketitle

\clearpage

\twocolumngrid

\textit{Introduction:}  Coherent elastic neutrino-nucleus scattering (\cevns{}) is a neutral current process~\cite{PhysRevD.9.1389,Kopeliovich:1974mv} with low momentum transfer, ($Q^2$), where the neutrino interacts coherently with the nucleus. The recoil energy transferred to the nucleus is observable, though typical recoil energies are low, tens of keV for neutrino energies in the tens of MeV range. Thus, detectors with low-energy thresholds are required for \cevns{} measurement.

\cevns{} has the largest cross section among neutrino scattering channels for $E_\nu<100$~MeV for most target nuclei.  The standard-model (SM) prediction depends on the nuclear weak charge, $Q_W^2$~$=$~$(N$$-$$(1$$-$$4\sin^2\theta_W)Z)^2\approx N^2$, where $N$ and $Z$ are the neutron and proton numbers of the target nucleus, and $\theta_W$ is the weak mixing angle \cite{Cadeddu:2018izq}.  CEvNS was first measured using the COHERENT CsI[Na] detector in an intense, pulsed source of neutrinos produced at the Spallation Neutron Source (SNS)~\cite{Mason:2000wb,Kustom:2000rj} at Oak Ridge National Laboratory~\cite{Akimov:2017ade}.



The COHERENT experiment deploys several detectors designed to measure \cevns{} and other low-energy scattering processes using the $\pi^+$ decay-at-rest ($\pi$DAR) neutrino flux at the SNS, attractive for \cevns{} measurements \cite{Scholberg:2005qs}. The detectors are situated in ``Neutrino Alley" (NA), a basement hallway where background neutrons from the facility are heavily suppressed. \cevns{} was first observed in NA, 19.3~m from the neutrino source using a 14.6~kg CsI[Na] scintillating detector \cite{Akimov:2017ade} 43 years after its theoretical prediction~\cite{PhysRevD.9.1389}. COHERENT also made the first detection of \cevns{} on argon \cite{Akimov:2020pdx}, which, together with the initial CsI[Na] measurement, agrees with the $N^2$ scaling of the cross section. While these campaigns were highly successful, they suffer from large statistical and systematic uncertainties, which limit their sensitivity to searches for new physical phenomena.





\cevns{} is a precisely predicted neutrino interaction within the SM.  The theoretical uncertainty is dominated by understanding of the spatial distribution of the weak charge in the nucleus. As a result, \cevns{} is a process well suited for probing physics beyond the SM (BSM).  A precision measurement of \cevns{} is sensitive to new particles, such as a dark photon that interferes with $Z$ exchange in the low-$Q^2$ regime \cite{Davoudiasl:2014kua,Liao:2017uzy,Miranda:2020tif} and may explain the g-2 anomaly \cite{Muong-2:2021ojo}.  Similarly, through the reliance of $Q_W^2$ on $\sin^2\theta_W$, \cevns{} may identify new physics through an unexpected value of the weak mixing angle at $Q^2\approx(50\text{ MeV})^2$ \cite{Miranda:2020tif}.  It can shed light on new forces at high mass scales through non-standard interactions (NSI) searches \cite{Barranco:2005yy}, the understanding of which is crucial for interpreting neutrino oscillation measurements, as NSI scenarios can obfuscate the interpretation of results \cite{Coloma:2019mbs,Denton:2020uda}.


Detectors that measure \cevns{} are also sensitive to sub-GeV, accelerator-produced dark matter particles~\cite{Dutta:2019eml,COHERENT:2019kwz}.  Further, \cevns{} from solar and atmospheric neutrinos are a background for dark matter direct detection experiments~\cite{Billard:2013qya,OHare:2016pjy,Boehm:2018sux}, making up the so-called neutrino floor, so that a clear understanding of their interaction will soon become paramount.  

\cevns{} will also contribute to measuring a future supernova neutrino burst~\cite{Horowitz:2003cz,Lang:2016zhv}.  As a neutral-current process, \cevns{} is sensitive to the total neutrino flux, which is of particular interest as other detection channels are most sensitive to the $\nu_e$~\cite{DUNE:2020jqi} or $\bar{\nu}_e$~\cite{Super-Kamiokande:2016kji} flux.  \cevns{} is also understood to play an important role in energy transport driving the core-collapse mechanism in the supernova~\cite{PhysRevLett.32.849,PhysRevLett.34.113,doi:10.1146/annurev.ns.27.120177.001123,Balasi:2015dba}.  



It is with precision measurements of \cevns{} that these physics searches are realized. In this letter, we present the first such measurement with the final CsI[Na] dataset and improved understanding of systematic uncertainties. Using the time structure of the neutrino flux from $\pi$DAR, leading constraints on non-standard neutrino interactions are presented, along with a direct measurement of the weak mixing angle at low $Q^2$.


\textit{Experiment:} We used a 14.6-kg scintillation CsI[Na] crystal \cite{Akimov:2017ade}.  The dopant was selected to reduce the rate of afterglow scintillation following a burst of activity in the detector \cite{Collar:2014lya}.  The crystal was attached to a single Hamamatsu R877-100 photomultiplier (PMT).  The signal was digitized at a rate of 500~MS/s with a dynamic range extending beyond the 60-keV$_{ee}$ calibration scale.  This crystal was shielded with both low-activity lead and low-$Z$ materials to mitigate $\gamma$ and neutron backgrounds~\cite{Collar:2014lya,2014PhDT.......121F}.  Muon veto panels surrounded the detector which allowed for removal of cosmic-associated activity.  


Our dataset includes 13.99~GWhr of integrated beam power that passes livetime criteria on beam stability, detector condition, and afterglow rate.  During data collection, the SNS ran using a mercury target with a mean beam energy of 0.984~GeV yielding $3.20\times10^{23}$ protons-on-target (POT).  Averaged over beam energies, a pion yield of $0.0848$~$\pm$~$10\%$~$\pi^+/$POT is expected from a G$\textsc{eant}$4~\cite{Agostinelli:2002hh} simulation of the SNS beam~\cite{Akimov:2021geg}.  The POT timing distribution averaged over the running period is calculated using beam current data from the SNS and has a FWHM of 378~ns.  Since this is less than the muon lifetime, the flux separates into two populations: a prompt, predominantly $\nu_\mu$ flux from $\pi^+$ decay followed by a delayed flux of $\nu_e$ and $\bar{\nu}_\mu$ from subsequent $\mu^+$ decay.  Over 99$\%$ of the SNS neutrino flux is generated by $\pi^+$ decay-at-rest~\cite{Akimov:2021geg}.


The detector was calibrated with the 59.5~keV $\gamma$ decay of an $^{241}$Am source.  With a Gaussian fit to calibration data, we found a light yield of 13.35 photoelectrons per keV electron-equivalent (PE/keV$_\text{ee}$).  Calibration data were taken with the source at nine different locations along the crystal, finding a spatial spread in light yield less than 3$\%$.  This is negligible compared to other identified energy smearing effects.  The single PE (SPE) charge was monitored during SNS running by tagging single PMT pulses with little other activity in the crystal.  

\textit{Data analysis:} Our analysis procedure closely parallels the approach described in \cite{Scholz:2017ldm,Akimov:2017ade} with improvements to our simulation, re-optimization of our event selection, and a more thorough detector response model.  Data coincident with the arrival of beam were blinded until reconstruction, selection, and analysis methods were determined.  Event time and energy were reconstructed by analyzing the PMT waveform in the beam window.  



The PMT voltage traces were digitized and a 70~$\mu$s waveform was saved for every beam spill.  We formed a 15~$\mu$s region-of-interest (ROI) coincident with the arrival of beam and formed a 3~$\mu$s integration time to capture most light given by a dominant scintillation decay constant $\tau=0.6$~$\mu$s \cite{Collar:2014lya}.  We also analyzed a 40~$\mu$s pretrace region (PT) immediately preceding the ROI which monitors afterglow activity in the crystal on a spill-by-spill basis.  We also analyze an analogous anti-coincident (AC) region preceding the beam to monitor steady-state backgrounds (SSBkg).

We applied two selection cuts to the waveform PT.  First, backgrounds producing afterglow contamination in the signal ROI are more likely to have more activity in the PT; we therefore only selected events with five or fewer PT pulses.  We also removed events that have a pulse within the last 200~ns of the PT which are typically background events that scatter very late in the PT and then leak into the ROI.  

Only events with $\geq9$ pulses reconstructed in the ROI are selected.  This mitigates background from coincidence of afterglow pulses.  These events are predicted to be biased to early scattering times in the ROI, with approximately exponential shape, $\tau\approx4$~$\mu$s.  Using this time dependence, we validated this simulation by comparing the rate and time dependence of the afterglow background using AC data and confirm that a negligible afterglow rate, consistent with 0, is expected after the $\geq9$~pulse cut.  This cut sets the analysis threshold, at $\approx7$~PE.



We applied nuclear recoil quenching by fitting the scintillation response curve, $E_{ee} = f(E_{nr})$, to five datasets collected in CsI[Na] including three taken by COHERENT~\cite{COHERENT:2021pcd,Collar:2019ihs}.  The recoil energies in these datasets spanned from 3 to 63~keVee.  To account for shape as a function of $E_{nr}$, we parameterized the scintillation response curve as a fourth degree polynomial, constrained so that $f(0)=0$.



The selection efficiency for \cevns{} recoils depends on observed energy, PE, and recoil time, $t_\text{rec}$.  We estimated energy dependence of the efficiency and its uncertainty using $^{133}$Ba calibration data which gave a sample of Compton-scattered electrons.  A coincidence with a backing detector was used to mitigate background and ensure only low-energy forward scattering events were used in the calibration.



There is a 39$\%$ chance that there is at least one afterglow pulse in each waveform ROI.  Since we reconstructed $t_\text{rec}$ as the time of the first pulse in the ROI, it is possible for a \cevns{} recoil occurring at late $t_\text{rec}$ to be rejected because it follows a random pulse which is accounted for in a time-dependent efficiency, $\varepsilon_T$, estimated with a data-driven simulation.  A library of waveforms from AC data was constructed by selecting exactly one waveform for each hour of detector running.  A simulated \cevns{} waveform was then overlaid on a waveform randomly selected from this library.  We took $\varepsilon_T$ as the ratio of events selected when simulated at $t=t_\text{rec}$ compared to $t=0$.  We also expect signal events that follow a random afterglow pulse but within the 3~$\mu$s integration window.  These events may be selected, but would have biased recoil energy and time.  This background was mitigated by requiring the time difference between the first and second pulse in the ROI be $<$ 520~ns.  This cut rejected a negligible fraction of events with properly reconstructed $t_\text{rec}$ but reduced the fraction of biased events sufficiently that the bias does not noticeably affect the measurement.  This was validated with large PE inelastic signals in our detector whose onset time was unambiguous.



Our energy resolution is dominated by photon counting.  However, the variation in SPE charge is also included in our energy resolution.  Combining these two effects, the smearing was modeled with a gamma function which appropriately predicts the asymmetric simulated smeared distribution much better than a Gaussian model.


Over 98$\%$ of the background comes from beam-uncorrelated, steady-state background (SSBkg).  This background is measured in-situ from AC data.  We estimated the PE distribution using all events found in AC data and used an exponential model for the time distribution with $\tau=20.2\pm2.6$~$\mu$s, consistent with the time dependence of the signal efficiency.  Uncertainty in this decay constant had a negligible impact on the measured cross section.


\begin{figure*}[!bt]
\centering
\includegraphics[width=0.49\textwidth]{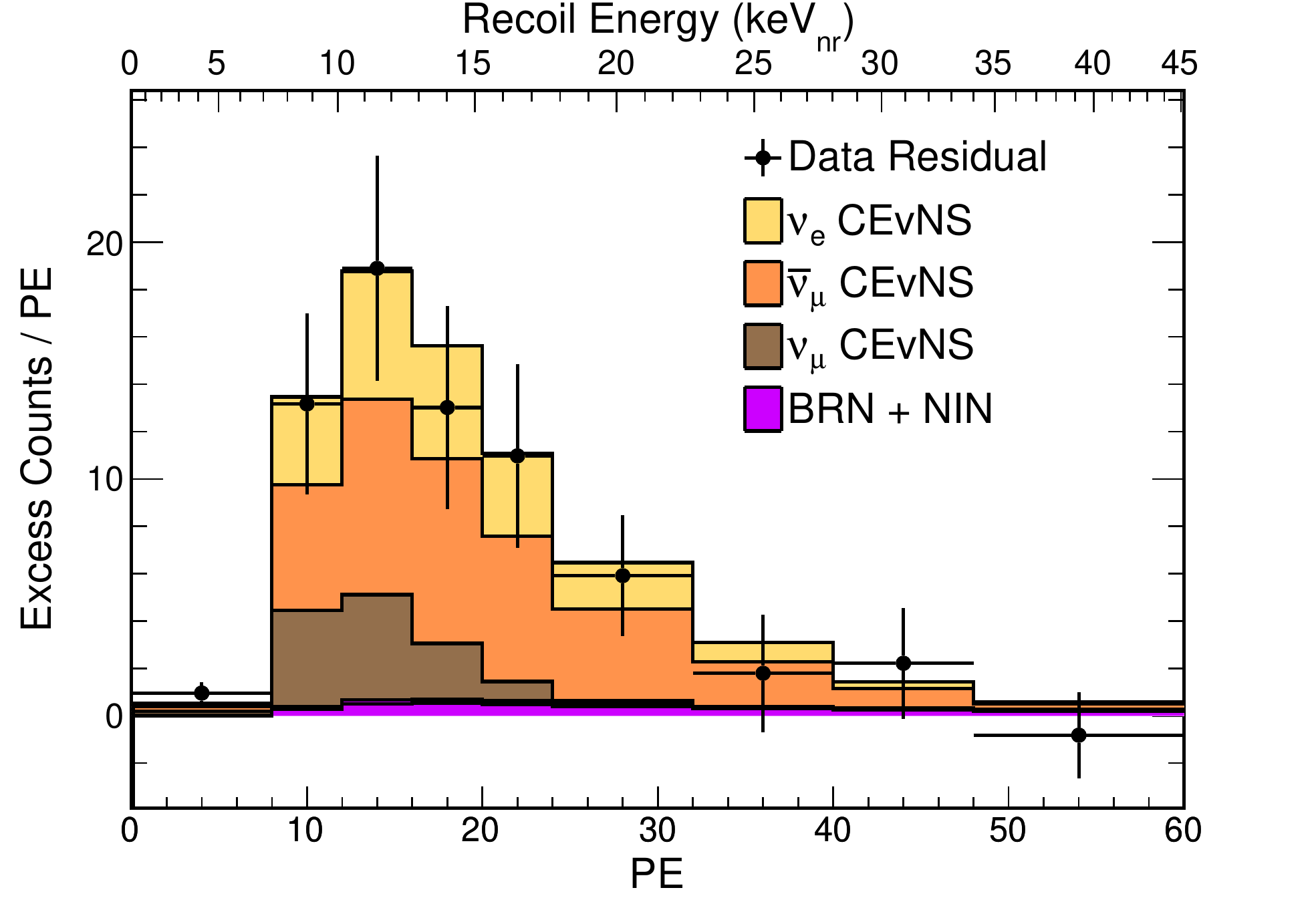}
\includegraphics[width=0.49\textwidth]{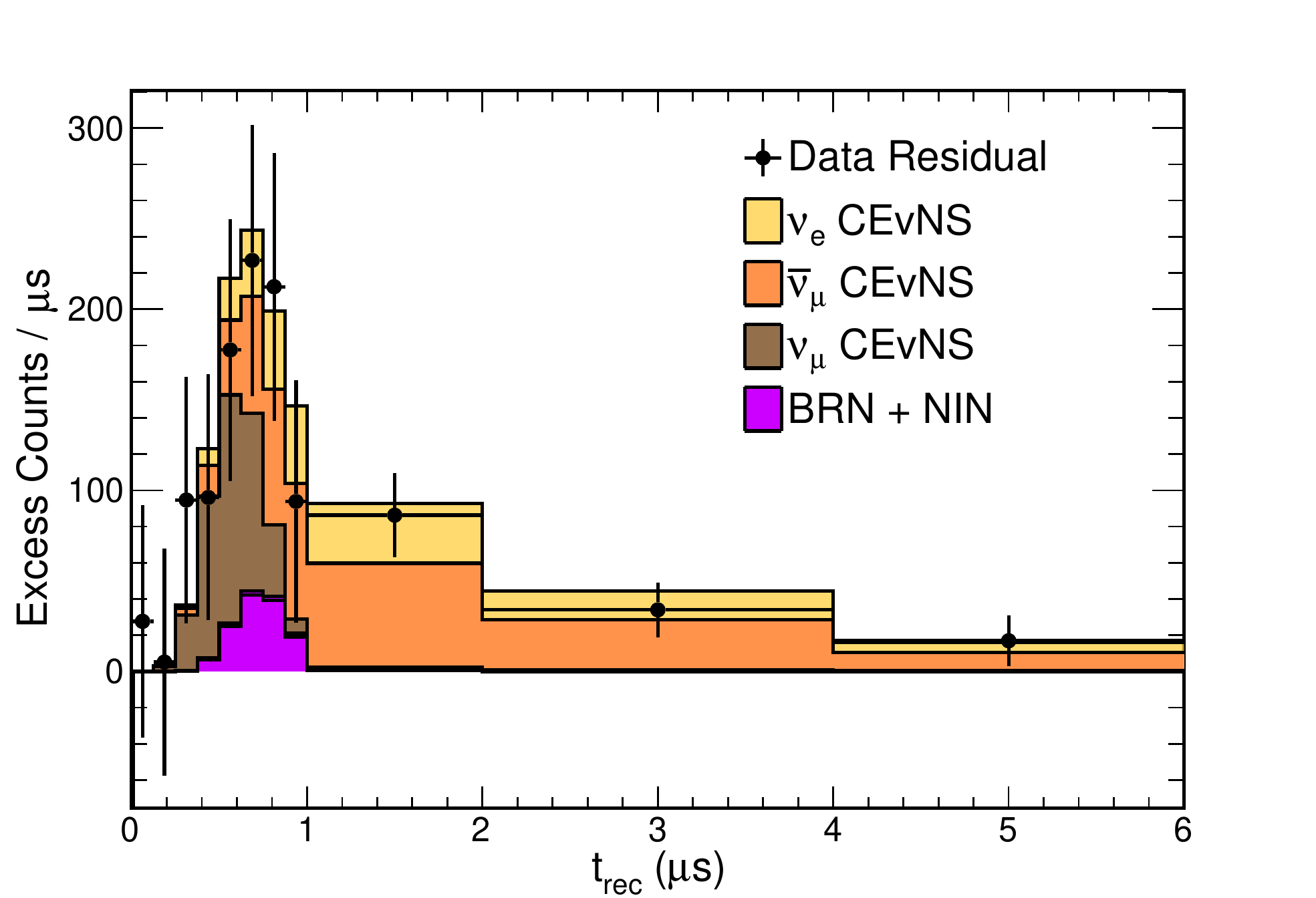}
\caption{The data residual over SSBkg background compared to best fit \cevns{}, BRN, and NIN predictions projected onto the PE (left) and $t_\text{rec}$ (right) axes.  The \cevns{} distribution has been decomposed into each flavor of neutrino flux at the SNS.}
\label{fig:ResidualCEvNS}
\end{figure*}

We accounted for two sources of beam-related background: beam-related neutron (BRN) and neutrino-induced neutron (NIN) scatters.  Prior to detector installation, the normalization of each of these components was studied by an EJ-301 liquid scintillator detector \footnote{Eljen Technology, 1300 W. Broadway St., Sweetwater, TX 79556} housed in the CsI[Na] shielding.  The neutron-moderating water used in the detector shielding was drained to increase the neutron rate.  The BRN and NIN rates were determined from a fit to the time distribution~\cite{Akimov:2017ade}.  A MCNPX-PoliMi~\cite{POZZI2003550} simulation was used to estimate the total flux of neutrons from each source incident on the EJ-301 detector.  This flux was then propagated through the full shielding into the CsI[Na] detector to simulate the neutron background.  We assume a power-law BRN flux, $\phi\propto E^{-\alpha}$.  Changes in the value of $\alpha$ have a negligible effect on the shape of our background distributions.  The NIN spectrum was estimated using MARLEY \cite{Gardiner2018,MARLEYv1.2.0} tuned to production on $^{208}$Pb with an incident $\pi$DAR spectrum.  After selection, we estimated $18\pm25\%$ BRN and $6\pm35\%$  NIN events in our sample with uncertainty dominated by the statistical precision of the EJ-301 fit \cite{Akimov:2017ade}.  Together BRN and NIN backgrounds are small, about 7$\%$ of the predicted \cevns{} rate.  



We performed a binned likelihood fit to data in both PE and $t_\text{rec}$.  All data events with $\text{PE}<60$ and $t_\text{rec}<6$~$\mu$s were included in the fit.  Systematic uncertainties were included as nuisance parameters including shape effects.  Uncertainty parameters were profiled in the fit.  We accounted for normalization uncertainty on each component.  The \cevns{} uncertainty is 10$\%$, dominated by the understanding of the total neutrino flux~\cite{Akimov:2021geg}.  We also included a 2.1$\%$ uncertainty on the SSBkg normalization due to a finite sample used to estimate the background.  



We also fit five systematic parameters that affect the shape of our predicted spectra.  The timing onset of the neutrino flux through our detector was allowed to float without any prior constraint.  Uncertainty in quenching was calculated by a principle component analysis (PCA) of the covariance matrix from fit to available data.  We identified two impactful uncertainties from the PCA giving a combined 3.8$\%$ bias in our fit.  A PCA was also performed on our \cevns{} efficiency curve from $^{133}$Ba calibration data.  This resulted in one systematic parameter which is roughly equivalent to a 1.0~PE uncertainty in threshold and gives a 4.1$\%$ uncertainty.  Finally, our form-factor uncertainty adjusts the neutron radius in CsI, $R_n$, by $\pm5\%$, which shifts the theoretical \cevns{} cross section by 3.4$\%$ and gives a 0.6$\%$ uncertainty on our measured cross section.  NSI scenarios would affect form-factor suppression~\cite{Hoferichter:2020osn}, but this effect has a negligble impact on constraints and is dropped.

\textit{Results:} After fitting, we observed $306\pm20$ \cevns{} events, consistent with the SM prediction of 341~$\pm$~$11(\textrm{theory})\pm42(\textrm{experiment})$.  The best-fit residual \cevns{} spectra in PE and $t_\text{rec}$ are shown in Fig.~\ref{fig:ResidualCEvNS}.  The best-fit prediction models the observed data well with a $\chi^2/\textrm{dof}=82.6/98$.  No excess is observed in beam-off data.  The cross section averaged over the $\nu_\mu/\nu_e/\bar{\nu}_\mu$ flux, $\langle\sigma\rangle_\Phi$, was determined to be $(165^{+30}_{-25})\times10^{-40}$~cm$^2$ by a profiled log-likelihood fit.  This is consistent with the SM prediction of $(189\pm6)\times10^{-40}$~cm$^2$.  The observed data reject the no-\cevns{} hypothesis at 11.6~$\sigma$.  See supplemental material at [\textit{URL}] to see observed data listed along with assumptions required to reproduce this result.





\begin{figure}[!bt]
\centering
\includegraphics[width=0.49\textwidth]{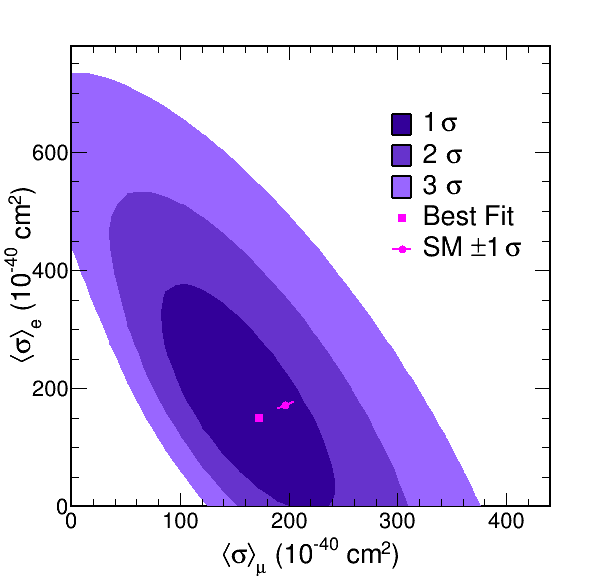}
\caption{Contours for the flavored \cevns{} cross section.  The best-fit parameters and the SM prediction, along with $\pm1$~$\sigma$ error bands from form-factor uncertainty, are shown as pink markers.}
\label{fig:FlavoredCEvNSXSec}
\end{figure}

Since the SM cross section depends on the weak charge, the \cevns{} cross section can be interpreted as a constraint on the weak mixing angle at a low momentum exchange, $Q^2\approx(50\text{ MeV})^2$ consistent with previous results \cite{PhysRevLett.111.141803}.  Our current result implies $\sin^2\theta_{W}=0.220^{+0.028}_{-0.026}$ compared to the SM prediction 0.23857(5) \cite{Zyla:2020zbs}.  Current constraints at low-$Q^2$ from atomic parity violation measurements are much more precise, though a percent-level measurement from COHERENT will be possible within the future~\cite{Akimov:2022oyb}.  Additionally, as $^{133}$Cs is a commonly used atom for these studies~\cite{Wood1759,PhysRevLett.109.203003}, \cevns{} data can be used to constrain theoretical uncertainties on nuclear structure assumed in these results \cite{Cadeddu:2018izq}.  


The ``flavored" \cevns{} cross sections, $\langle\sigma\rangle_\mu$ and $\langle\sigma\rangle_e$ are also measured by exploiting the differences in timing shapes between the \cevns{} contributions from $\nu_\mu$, $\bar{\nu}_\mu$ and $\nu_e$.  This parameter space is a sensitive probe of BSM physics such as neutrino-quark vector NSI which can affect each neutrino flavor differently \cite{Barranco:2005yy}.  The flavored \cevns{} cross section result is uniquely possible using a flux from a spallation sources with beam width less than the muon half-life.  The allowed contour in this parameter space is shown in Fig.~\ref{fig:FlavoredCEvNSXSec}.  The best-fit scales relative to the SM are 0.88 and 0.87 for $\langle\sigma\rangle_\mu$ and $\langle\sigma\rangle_e$, respectively, consistent with the SM.





\begin{figure}[!bt]
\centering
\includegraphics[width=0.49\textwidth]{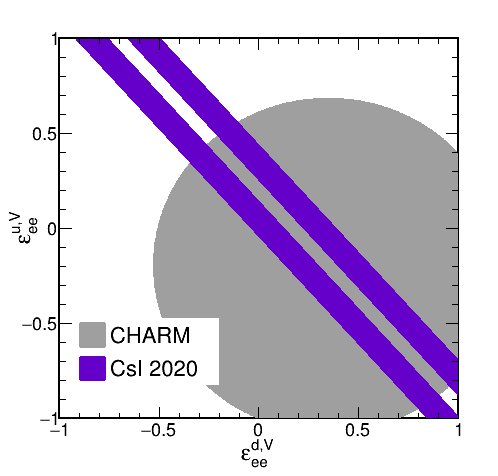}
\includegraphics[width=0.49\textwidth]{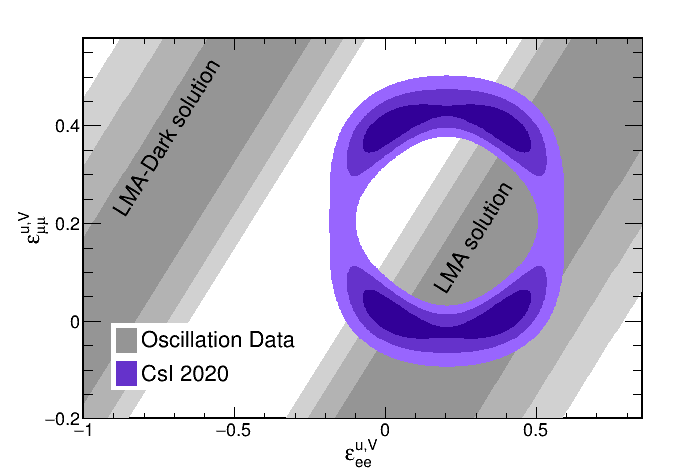}
\caption{The top plot shows the 90$\%$ allowed parameter space with $\varepsilon_{ee}^u$ and $\varepsilon_{ee}^d$ to float while fixing others at 0, while the bottom shows 1/2/3$\sigma$ contours allowing $\varepsilon_{ee}^u$ and $\varepsilon_{\mu\mu}^u$ to float fixing others to 0.  The bottom also shows parameter space that is compatible with a degeneracy in solar neutrino oscillation data that would flip the inferred neutrino mass ordering.}
\label{fig:NSIConstraints}
\end{figure}

We used this measurement to constrain heavy-mediator ($m_V\gg Q$) neutrino-quark NSI, commonly parameterized as a matrix of $\varepsilon_{ij}^f$ where $i,j=e,\mu,\tau$ and $f=u,d$.  Existence of NSI could confuse ongoing efforts to measure the neutrino mixing matrix parameters.  Notably, it is possible to reverse the inferred neutrino mass ordering from oscillation data by choosing a suitable set of NSI parameters \cite{Coloma:2019mbs}.  Also, NSI allow for additional $CP$-violating phases which may bias constraints on $\delta_{CP}$ \cite{Denton:2020uda,Khan:2021wzy}.


In Fig.~\ref{fig:NSIConstraints}, we show the constraint on $\varepsilon_{ee}^u$ and $\varepsilon_{ee}^d$ with other parameters fixed to 0 compared to CHARM \cite{DORENBOSCH1986303} constraints.  This marks a significant improvement over the previous CsI[Na] constraint from COHERENT~\cite{Akimov:2017ade} because of an improved precision result and measuring the flavored cross sections.  There are also NSI constraints determined from \cevns{} data on Ar \cite{Akimov:2020pdx} and Xe \cite{Aprile:2020thb}, though these limits are currently less precise.

Fig.~\ref{fig:NSIConstraints} also shows our sensitivity to $\varepsilon_{ee}^u$ and $\varepsilon_{\mu\mu}^u$.  This combination is directly related to solar neutrino oscillation results.  In the context of NSI, there is a degeneracy in oscillation data between the large mixing angle (LMA) and LMA-Dark solutions which differ in the $\theta_{12}$ octant and altering the interpretation of the neutrino mass ordering \cite{Coloma:2017ncl}.  The shape of the allowed parameter space again highlights the power of the flavored \cevns{} measurement as $\varepsilon_{ee}^{u,V}$ and $\varepsilon_{\mu\mu}^{u,V}$ only affect the \cevns{} cross section for $\nu_e$- and $\nu_\mu$-flavor neutrinos, respectively.


\textit{Conclusion:} We measured the \cevns{} cross section using the full dataset collected by the CsI[Na] scintillation detector using a blinded analysis approach.  With doubled exposure and improved understanding of systematic uncertainties, we have made the most precise measurement of \cevns{} to date, observing \cevns{} at 11.6~$\sigma$ and finding a flux-averaged cross section $\langle\sigma\rangle_\Phi=(165^{+30}_{-25})\times10^{-40}$~cm$^2$, consistent with the SM prediction to within 1~$\sigma$.  The weak mixing angle was measured at low $Q^2$.  We also introduced measurements of the flavored \cevns{} cross section, which improve \cevns{} constraints on neutrino-quark NSI scenarios.  Though the CsI[Na] detector has been decommissioned, a planned calibration of the neutrino flux using a heavy-water Cherenkov detector~\cite{Akimov_2021nkt} will further improve precision of the \cevns{} measurements.  COHERENT is currently engaged in ongoing measurements of \cevns{} on Ar, Ge, and NaI, while additional targets are possible for the future.


\textit{Acknowledgements:} The COHERENT collaboration acknowledges the Kavli Institute at the University of Chicago for CsI[Na] detector contributions.  The COHERENT collaboration acknowledges the generous resources provided by the ORNL Spallation Neutron Source, a DOE Office of Science User Facility, and thanks Fermilab for the continuing loan of the CENNS-10 detector. We also acknowledge support from the Alfred~P. Sloan Foundation, the Consortium for Nonproliferation Enabling Capabilities, the National Science Foundation,  the Russian Foundation for Basic Research (proj.\# 17-02-01077 A), and the U.S. Department of Energy, Office of Science. Laboratory Directed Research and Development funds from ORNL and Lawrence Livermore National Laboratory also supported this project.  This research used the Oak Ridge Leadership Computing Facility, which is a DOE Office of Science User Facility.  Sandia National Laboratories is a multi-mission laboratory managed and operated by National Technology and Engineering Solutions of Sandia LLC, a wholly owned subsidiary of Honeywell International Inc., for the U.S. Department of Energy’s National Nuclear Security Administration under contract DE-NA0003525.  The work was supported by the Ministry of Science and Higher Education of the Russian Federation, Project Fundamental properties of elementary particles and cosmology No. 0723-2020-0041

%

\end{document}